\documentclass[prd,aps,twocolumn,showpacs]{revtex4}
\usepackage{epsfig}
\begin{document}
\title{Localized gravitational energy in a Schwarzschild field}

\author{J Kentosh}

\email{james.kentosh.336@my.csun.edu}

\affiliation{Department of Physics and Astronomy, California State University,
 Northridge, California, USA}

\date{November 27, 2012}

\pagenumbering{arabic}

\begin{abstract}
An interpretation of general relativity is developed in which the energy used to lift a body in a static gravitational field increases its rest mass. Observers at different gravitational potentials would experience different mass reference frames.  It is shown that bodies falling in a Schwarzschild field exhibit the relativistic mass/velocity relationship from special relativity.  This new result is independent of the choice of coordinates.  The proposed approach provides a physical explanation for gravitational energy, which is localized as a scalar function intrinsic to general relativity.  Applying this model to the Robertson-Walker metric demonstrates that time-varying fields induce a net energy transfer between bodies that is not exhibited in static fields.
\end{abstract}

\maketitle

\section{Introduction}
Over 30 years ago Rafael Vera introduced fascinating concepts for mass \cite{vera81}.  They have not gained widespread acceptance, possibly because they were thought to be incompatible with general relativity (GR).  In this paper, a careful accounting of observer reference frames shows that Vera's ideas are consistent with GR in a Schwarzschild field.  This combination is an attempt to begin to resolve the long-standing perplexity over the localization of gravitational energy.

One of Vera's ideas is that the rest mass of a particle increases with its gravitational potential \cite{vera81}.  Therefore, gravitational energy would be localized with the masses of bodies.  This must occur in such a way that proper (locally measured) rest mass is invariant with position, as perceived from a local observer's mass reference frame.  The mass/velocity relationship for a body falling along a geodesic in a Schwarzschild field is shown to be the old-fashioned expression for relativistic mass from special relativity.  This new result is independent of the choice of coordinates, as confirmed for the isotropic and harmonic metrics.  This approach strengthens the correspondence between special relativity and GR, and rehabilitates the undervalued concept of relativistic mass.

Reference frames are usually associated with spatial coordinates and time.  The introduction of reference frames for mass is essential for solving this problem.

Applying Vera's concepts to time-varying gravitational fields is problematic.  It is shown that his concept of an invariant falling mass does not apply to such fields.  Unlike static fields, time-varying fields induce a net long-range energy transfer between bodies.

The paper proceeds with the following steps:  A dimensionless scalar $\beta_{ab}$ is introduced; its important identities are found.  A mass transformation is derived for a one-dimensional, static gravitational field, then generalized to transformations between mass reference frames.  The transformations are applied to the Schwarzschild solution and other metrics. The mass/velocity relationship for falling bodies is evaluated for radial and oblique motion. It is shown that the proposed interpretation conserves energy and mass within a Schwarzschild field while the conventional interpretation does not.  Time-varying fields are studied and limitations on the proposed interpretation are imposed.  As with Maxwell's equations, in which a time-varying electric field causes effects not seen in static fields, a time-varying gravitational field also causes effects not exhibited by static fields. This work is a matter of interpretation and is intended to complement GR.

\section{Background}
What happens to the energy used to lift a body in a gravitational field?   During the development of GR, Einstein derived an expression $t^{\mu\nu}$ for the stress-energy of the gravitational field.  That expression is a pseudotensor that varies with the choice of coordinates in an unseemly way. (For a review see \cite{szabados09, baryshev08}.)  Many attempts have been made to better understand gravitational energy in GR.  Tolman \cite{tol30} defined gravitational energy as a separate term in the energy-momentum tensor $T^{\mu \nu}$.  In 1935, Whittaker \cite{whit35} proposed the term ``potential mass" to describe the contribution of gravitational potential energy to the gravitating mass of a particle.  After much deliberation, the interpretation arose that gravitational energy is not localized in GR \cite{landau59, dirac75}. Thus, it is meaningless to ask where gravitational energy resides.  In an influential textbook \cite{mtw72}, Misner, Thorne and Wheeler sought to resolve this quandary with the words, ``Anybody who looks for a magic formula for `local gravitational energy-momentum' is looking for the right answer to the wrong question.  Unhappily, enormous time and effort were devoted in the past to trying to `answer this question' before investigators realized the futility of the enterprise." Yet Bondi insists that non-localizable forms of energy are inadmissible in GR \cite{bondi90}.

Although today most relativists seem satisfied with the conventional treatment of gravitational energy as unlocalizable, there remains considerable ambiguity in the interpretation of gravitational energy within GR.  Many authors believe that gravitational energy, and even mass, are stored in fields \cite{ohanruff94, steph90, pauli58}.  Some conclude that gravitational self-energy, which is a type of gravitational energy, contributes to the masses of bodies \cite{baessler99, uzan03}.  Vera \cite{vera81}, Savickas \cite{savickas94}, and Ghose and Kumar \cite{ghosekum76} interpret rest mass as depending on local gravitational potential, while Ohanian and Ruffini \cite{ohanruff94} argue that rest mass is a constant independent of space and time. Since energy cannot simultaneously be distributed in fields and contribute to particle masses, some of these views appear to be incompatible.  Hayward states, ``It comes as a surprise to many that there is no agreed definition of gravitational energy (or mass) in general relativity" \cite{hayward94}.

A few dedicated researchers continue to study gravitational energy and related concepts such as quasi-local energy momentum (e.g., \cite{szabados09, chang99, wu11, wang11, mitra10, zhang09, liu03, frauendiener11, katz05}), driven partly by the belief, as stated by Mirshekari and Abbassi \cite{mirshekari08}, that ``one of the old and basic problems in general relativity which is still unsolved is the localization of energy."

\section{Concept for gravitational energy}
As proposed or implied by others (e.g., \cite{vera81, savickas94, ghosekum76}) it is hypothesized that the energy used to lift a body in a static gravitational field increases its rest mass.  It is also assumed that the mass of a body remains constant while falling in a static field \cite{vera81}.  This interpretation is illustrated in Fig. 1 with the following steps:  The energy $\Delta E$ used to lift a body increases its rest mass by $\Delta E/c^2$.  When the lifted body is allowed to fall from a higher elevation, it maintains a constant mass during its free-fall, equal to its mass at the point it began its fall.  As the body passes its starting point with some velocity, it has a greater mass than an identical body at rest there, in general agreement with special relativity.  When the fall is stopped, the body loses its extra mass and returns to the rest mass it had before being lifted.  The mass that the body loses during its impact is transferred to other bodies (and to itself) in the form of vibrational energy, heat, elastic or inelastic energy, etc.  Total mass would always be conserved during lifting, falling, and collisions.  No mass or energy would be stored in the gravitational field.

\begin{figure}[htb]
\begin{center}
\epsfig{file=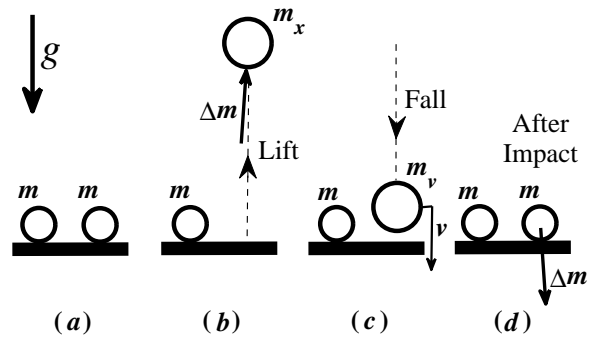, width=8.5 cm, angle = 0}
\caption{({\it a}) Two identical bodies with mass $m$ at rest in a static gravitational field.  ({\it b}) One is lifted and the lift energy increases its mass by $\Delta m$ to $m_x$.  ({\it c}) The lifted body falls back to its starting point.  While falling it maintains a constant mass of $m_v=m_x$ and has a greater mass than the body at rest.  ({\it d}) The fall is stopped and the body returns to its original mass $m$.  Its extra mass $\Delta m$ is transferred to other bodies during the collision.}
\end{center}
\end{figure}

With this interpretation, the rest masses of all bodies would depend on position in a gravitational field.  In accordance with the equivalence principle, any experiments conducted at different elevations should be unable to detect any difference in mass, since all objects would be affected in exactly the same proportion.  This interpretation does for potential energy what special relativity did for kinetic energy \--- gives it a mass equivalent.  The history of various concepts of gravitational energy is a fascinating topic not covered herein.

The concepts shown in Fig. 1 appear to work in static fields. However, it is shown later that the mass of a body cannot remain constant while falling in a \emph{time-varying} field, in which gravitational energy must be conveyed between bodies via gravitons and/or gravitational waves. The essential element of Fig. 1 that remains applicable in time-varying fields is that energy and mass are primarily localized as part of the masses of bodies rather than distributed in fields.

For the purpose of this paper, no distinction needs to be made between inertial mass, active gravitational mass, and passive gravitational mass.  They are assumed to be the same \cite{rosen92, roche05}.  Also, the term ``gravitational energy" refers to the gravitational potential energy of a body and is not to be confused with the more general gravitational energy-momentum tensor $T^{\mu \nu}$.

As an example of the conservation of mass after an inelastic collision, thermal energy increases the velocity of the molecules of a body, with corresponding relativistic mass given by \begin{equation}\label{mKE}
m_v=\frac{m_o}{\sqrt{1-v^2/c^2}} \approx m_o + \frac{m_o v^2/2}{c^2} = m_o + \frac{KE}{c^2},
\end{equation} so that the mass of a falling body is in part converted into the mass associated with additional kinetic energy $KE$ of the molecules \cite{carlip98}.  After the body cools off, it returns to its original mass.

Although the terms ``rest mass" and ``relativistic mass" have fallen out of favor \cite{okun98, adler87, okun09}, they are used herein out of necessity.  Here the term ``rest mass" describes the mass of a motionless body in any reference frame fixed to a central mass in a static gravitational field.  Since reference frames span different gravitational potentials, rest masses may vary with elevation.  ``Proper rest mass" is the rest mass of a body when it is co-located with the observer measuring it; proper rest mass is presumed to be invariant.  ``Relativistic mass" refers to the mass of a body moving relative to an observer fixed with respect to the central mass, as perceived from that observer's mass reference frame. Rather than dwell on semantic arguments over the acceptability of the term ``relativistic mass," it is more productive to focus on physical arguments and their consequences.

MTW \cite{mtw72} derived localized gravitational energy for the Schwarzschild solution.  However, their approach differs from Vera's concepts \cite{vera81}.

Several variable mass theories of gravity have been proposed \cite{dicke62, malin74, hoyle75, bekenstein77}.  However, they do not directly relate changes in mass with potential energy.

The standard model extension (SME) has been developed partly to study possible variations in fundamental constants \cite{colladay97, colladay98, kostelecky04}.  In the SME, the relative masses of different materials might vary with gravitational potential, in disagreement with GR. That possibility is not considered herein.  Modern versions of the E$\ddot{\textnormal{o}}$tv$\ddot{\textnormal{o}}$s experiments \cite{dent08, wagner12} support the hypothesis that different materials are affected by gravitational fields in the same way.

\section{Dimensionless scalar $\beta$}
Because it is necessary to understand static fields before exploring time-varying fields, this work begins with a detailed analysis of the Schwarzschild solution.  For static gravitational fields it is useful to define the scalar dimensionless ratio $\beta_{ab}$ as follows: \begin{equation}\label{betadef}
\beta_{ab} \equiv \frac{dt_a}{dt_b},
\end{equation} where $dt_a$ is the time interval between two events measured at some point $A$ and $dt_b$ is the time interval between those two events measured from some other point $B$, using clocks that are motionless relative to the field.  If $\beta_{ab}<1$, time passes more slowly at point $A$ than at $B$.  (In a static field it is easy to account for the time for a signal to travel between $A$ and $B$.)

From the form of (\ref{betadef}) the following identities for $\beta$ can be found: \begin{equation}\label{bid1}
\beta_{aa}=\beta_{bb}=1, \end{equation}

\begin{equation}\label{bid2}
\beta_{ab}=\frac{1}{\beta_{ba}},
\end{equation} and \begin{equation}\label{bid3}
\beta_{ab} \beta_{bc}=\beta_{ac}.
\end{equation}

\section{General mass transformation}

The next step is to derive a general transformation for rest mass in a static, one-dimensional gravitational field,  from the perspective of observers who are motionless relative to the field.  Based on Einstein's equivalence principle, an observer at any location should measure the same rest mass for a given body when it is at her location.  This could be achieved experimentally by comparing the mass of the body to the mass of a kilogram standard carried alongside.

Consider an observer located at some arbitrary reference elevation $O$.  Let the proper rest mass of some body be $m_o$.  The mass of that body when it is at some other elevation $x$, relative to the reference frame at $O$, is defined to be $m_{xo}$ \--- an important notation for what follows.  The difficulties of measuring mass at another location must be overlooked for now. When the body is at $O$, the observer there will measure its mass to be $m_{oo}\equiv m_o$.  When the body is carried to some other elevation $x$, an observer there will measure its mass to be $m_{xx} = m_o$.

Consider a static field causing a proper acceleration $g(x)$ in the $-x$ direction.  Let $x$ represent proper distance, which will generally differ from the coordinates used in some solution to the field equations.  Consider a body with proper rest mass $m_o$ being slowly lifted against the field.  Then the proper energy required to lift the body some small distance $dx$ is given by \begin{equation}\label{de}
dE=m_o g(x) dx.
\end{equation}

In the local reference frame at $x$, the small increase in mass as the body is lifted is given by \begin{equation}\label{dmxx}
dm_{xx}=\frac{d E}{c^2}=\frac{m_o g(x) d x}{c^2}.
\end{equation}

From the reference frame at $O$, the body at $x$ has mass $m_{xo}$, while an observer at $x$ measures the mass of the body to be $m_o$.  The ratio of those masses will apply to any body or mass increment at $x$.  Therefore, the incremental mass $d m$ at $x$ observed from $O$ and from $x$ are related as follows: \begin{equation}\label{mratios}
\frac{d m_{xo}}{d m_{xx}}=\frac{m_{xo}}{m_{xx}}=\frac{m_{xo}}{m_o}.
\end{equation}

Combining (\ref{dmxx}) and (\ref{mratios}) yields \begin{equation}\label{dmxo}
\frac{d m_{xo}}{m_{xo}}=\frac{g(x)}{c^2} d x.
\end{equation}

This can be integrated to yield the general mass transformation between elevations in a one-dimensional, static gravitational field: \begin{equation}\label{mxox}
m_{xo}(x)=m_o \exp \left( \frac{1}{c^2} \int_{o}^{x} g(x) d x \right).
\end{equation} This equation, also found in \cite{savickas94}, indicates that the mass of a body lifted against the field to $x$ is greater than the mass of an identical body at $O$. Since there is no preferred elevation in a gravitational field, point $O$ can be selected to represent any point.  The equation also shows that proper mass does not vary with elevation.

In a weak field, (\ref{mxox}) reduces to $m_{xo} \approx m_o + (\Phi_x-\Phi_o)/c^2$, where $\Phi_i$ is the classical gravitational potential energy of the body.

Next the general transformations between mass reference frames can be found.  Consider the general case of two elevations $A$ and $B$.  Let $O \to A$ and $x \to B$.  From the form of the preceding equation, it can be shown that any rest mass $m_o$ at $B$ relative to the mass reference frame at $A$, and vice-versa, satisfies identities that are similar to those for $\beta$, so that \begin{equation}\label{mid1}
\frac{m_{aa}}{m_o} = \frac{m_{bb}}{m_o} = 1,
\end{equation} \begin{equation}\label{mid2}
\frac{m_{ab}}{m_o}= \left( \frac{m_{ba}}{m_{o}}\right)^{-1},
\end{equation} and \begin{equation}\label{mid3}
\frac{m_{ab}}{m_o} \frac{m_{bc}}{m_o}=\frac{m_{ac}}{m_o},
\end{equation} where $m_{ab}$, for example, is the mass of a body at $A$, with proper rest mass $m_o$, relative to the mass reference frame at $B$.  Observers at different gravitational potentials experience different mass reference frames that are related by these formulas.

\section{Mass transformation of the Schwarzschild solution}

Next, the mass transformation is applied to the exterior Schwarzschild solution of GR, described by the following metric: \begin{equation}\label{ds2}
d s^2=c^2\beta_{r\infty}^2 d t^2-\frac{d r^2}{\beta_{r\infty}^2}-r^2 d \theta^2 - r^2 \sin^2 \theta d \phi^2,
\end{equation} where \begin{equation}\label{brinf}
\beta_{r\infty}=\sqrt{1-2GM/c^2 r}.
\end{equation}

The coordinate time $d t$ is the time interval of a stationary clock at infinity and $d s = c \thinspace d \tau$, where $d \tau$ is proper time of a clock corresponding to the spacetime interval $d s$.  Setting $d r=d \theta = d \phi = 0$ in (\ref{ds2}) for a motionless clock at $r$ yields $\beta_{r\infty} = d t_r/d t_{\infty}$, where $d \tau \to d t_r$ is the time interval of a stationary clock at $r$ and $d t_{\infty}=d t$ is the time of a stationary clock at infinity.  Thus, $\beta_{r\infty}$ is the ratio of the time rates at $r$ and infinity, corresponding to the definition for $\beta$ in (\ref{betadef}).  That is the origin of the subscripts in (\ref{brinf}).

Equation (\ref{brinf}) provides $\beta$ at $r$ relative to an observer at infinity.  At some other radius $R$, \begin{equation}\label{bRinf}
\beta_{R\infty}=\sqrt{1-2GM/c^2 R}.
\end{equation}

From identities (\ref{bid2}) and (\ref{bid3}) an expression for the $\beta$'s relating elevations $r$ and $R$ can be found: \begin{equation}\label{brR}
\beta_{Rr} = \frac{\beta_{R\infty}}{\beta_{r\infty}}= \sqrt{\frac{1-2GM/c^2 R}{1-2GM/c^2 r}}.
\end{equation} Besides using the identities, this can also be derived from the relative time rates at $r$ and $R$.

Since the mass transformation (\ref{mxox}) is based on proper acceleration $g(x)$, the next step is to find proper acceleration relative to a motionless observer in a Schwarzschild field.  For radial motion, Hobson \emph{et al.} \cite{hobson07} find the following geodesic equation of motion: \begin{equation}\label{rddot}
\ddot{r} = - \frac{GM}{r^2},
\end{equation} where $\dot{r} \equiv d r/d \tau$.  In the general case, $\tau$ is the time measured by a clock that may be either moving or stationary. In the case of a stationary clock at $r$, the terminology $\tau \to t_r$ is used to differentiate between those two cases.  For a motionless clock at $r$, $d t/d \tau \equiv d t_{\infty}/d t_r =1/\beta_{r\infty}$.

Coordinate $r$ in the Schwarzschild metric has little physical significance.  Of greater relevance is the proper distance given by $d s$, which in the general case may correspond to a moving meter stick or one that is motionless relative to the central mass. To distinguish between those two possibilities, variable $x$ is used to denote proper distance in the reference frame of a motionless observer at $r$.  By setting $d t=d \theta =d \phi = 0$, letting $d s \to d x$ in (\ref{ds2}), and using the appropriate sign for a spacelike spacetime interval, the following relationship is found: \begin{equation}\label{drdx}
d r=\beta_{r\infty} d x,
\end{equation} which relates proper distance $d x$ to the radial coordinate distance $d r$.  Using the resulting formula for $d x$ in place of $d r$ in (\ref{rddot}), \begin{equation}\label{drdot}
\ddot{r}=\frac{d}{d \tau} \left( \beta_{r\infty} \frac{d x}{d \tau} \right) = \frac{d \beta_{r\infty}}{d\tau}
\frac{d x}{d \tau} + \beta_{r\infty} \frac{d^2 x}{d \tau^2}.
\end{equation}

Consider a body at the instant it is released to free-fall, before it has attained appreciable velocity.  Then $d \tau \approx d t_r$, the proper velocity is $d x/d \tau \approx 0$, and only the right hand term in (\ref{drdot}) contributes. Using the resulting equation in combination with (\ref{rddot}) and (\ref{drdx}), the proper acceleration of gravity for a slowly falling body at $r$ in a Schwarzschild field is given by \begin{equation}\label{gr}
g(r) \equiv  -\frac{d^2 x}{d t_r^2} \approx \frac{GM}{\beta_{r\infty} r^2}.
\end{equation} This is the acceleration that must be resisted when slowly lifting a body in the field.  Note that $g(r)$ is positive towards the central mass.

Suppose a small body with proper rest mass $m_o$ is lifted radially from $r$ to $R$ in a Schwarzschild field. The mass of the body at $R$, relative to the reference frame at $r$, can be found by substituting (\ref{drdx}) and (\ref{gr}) into (\ref{mxox}), which yields \begin{equation}\label{mRrexp}
m_{Rr}=m_o \exp \left( \frac{GM}{c^2} \int_{r}^{R} \frac{d r}{\beta_{r\infty}^2 r^2}\right).
\end{equation}

The surprisingly simple solution to this integral is \begin{equation}\label{mRrb}
m_{Rr} = \frac{\beta_{R\infty}}{\beta_{r\infty}} m_o = \beta_{Rr} m_o.
\end{equation} Since $\beta_{Rr}>1$ if $R>r$, the mass of the lifted body at $R$ is perceived to be greater than $m_o$ from the reference frame at $r$.  In a weak field, $m_{Rr}$ is its rest mass plus (or minus) the mass equivalent of the change in classical potential energy.

If $r$ is chosen to be the reference frame at an infinite distance from the central mass and if $R\to r$, then the relative mass of a body at any point $r$ perceived from the reference frame at infinity is given by \begin{equation}\label{mrinf}
m_{r\infty}= \beta_{r\infty} m_o.
\end{equation} Thus, stationary bodies in the vicinity of a central mass have a decreased mass relative to a reference frame at infinity.  It is apparent that $\beta_{r\infty}$ plays the role of scale factor for mass as well as time.

The preceding results for the Schwarzschild solution suggest the following general transformation for rest mass in any static gravitational field: \begin{equation}\label{mxo}
m_{xo}=\beta_{xo} m_o.
\end{equation} This generalization, offered without proof, does not depend on the choice of coordinates;  it is not used again in this paper. The proof of (\ref{mxo}) for any static field $g(x)$ requires topics not covered here.

Most studies of localization focus on the gravitating mass rather than test particles.  Nevertheless, mass reference frames reveal an interesting property of the gravitating mass in a Schwarzschild field. In some ways, the Schwarzschild metric is written from the perspective of a reference frame at infinity:  Coordinate time $t$ represents a clock at infinity and radial distance $r$ converges to flat space coordinates at infinity.  Therefore, the central mass at the origin should be $M \equiv M_{o \infty}$, its mass from the reference frame at infinity.  From a reference frame at $r$, all masses appear increased from (\ref{mRrb}), so that $M_{or}=M_{o\infty}/\beta_{r \infty}$.  Substituting this into (\ref{gr}) yields \begin{equation}\label{grr}
g(r)=\frac{G M_{or}}{r^2},
\end{equation} where $r$ is the proper circumference divided by $2\pi$.  Thus, Newton's law of gravity is preserved for any local observer if the magnitude of the central mass in his reference frame is used.  This result is independent of the choice of coordinates.

\section{Relativistic mass of a falling body}

Now that the mass transformation has been derived for a Schwarzschild field, the mass/velocity relationship for a falling body can be found.  Equation (\ref{mRrb}) provides the mass of a body that has been lifted from $r$ to $R$.  Based on the concept in Fig. 1, if that body is allowed to fall from $R$ back to $r$, it will maintain the same mass during its fall.  It will therefore arrive at $r$ with the relativistic mass $m_v = m_{Rr}$ given by (\ref{mRrb}).

In GR it is straightforward to derive the velocity of a body falling radially from $R$ to $r$ in a Schwarzschild field. The radial geodesic equations of motion \cite{hobson07} are \begin{equation}\label{rdot2}
\dot{r}^2 = c^2(k^2-1) + 2GM/r,
\end{equation} and \begin{equation}\label{tdot}
\left( 1-\frac{2GM}{c^2 r}\right) \dot{t} = k.
\end{equation}

Expressing $\dot{r}$ in terms of $d r/d t$ and setting constant $k$ so that the coordinate velocity $d r/d t=0$ at $R$, the coordinate velocity of the body when it arrives at $r$ is \begin{equation}\label{drdt}
\frac{d r}{d t}= c \beta_{r\infty}^2 \sqrt{1-\frac{\beta_{r\infty}^2}{\beta_{R\infty}^2}}.
\end{equation}

Using prior formulas, the proper velocity $v_r$ of a body at $r$ can be related to its coordinate velocity as follows: \begin{equation}\label{vrdrdt}
v_r \equiv \frac{d x}{d t_r} = \frac{1}{\beta_{r\infty}^2} \frac{d r}{d t},
\end{equation} where $v_r$ is measured with respect to proper coordinates $x$ fixed relative to the central mass.  The preceding two equations can be combined to yield the proper velocity of a body falling from $R$, measured at $r$ by a motionless observer there: \begin{equation}\label{vrsqrt}
v_r=c \sqrt{1- \frac{\beta_{r\infty}^2}{\beta_{R\infty}^2}} = c \sqrt{1- \beta_{Rr}^{-2}}.
\end{equation}

Using (\ref{mRrb}) to provide the locally measured mass of the falling body $m_v = m_{Rr}$, it is found that the relationship between proper velocity and total mass of the falling body at $r$ is given by \begin{equation}\label{mv}
m_v= \frac{m_o}{\sqrt{1-v_r^2/c^2}}.
\end{equation} This is the well known relativistic mass/velocity relationship of special relativity, derived for a body falling in a Schwarzschild field.

\section{Other initial conditions}

The mass/velocity relationship of (\ref{mv}) was derived for a body released from rest at $R$.  What if some initial, proper radial velocity $v_R$ is imparted to the body at $R$ as it begins its fall?  Then an observer at $R$ will perceive the body to have an increased mass as it begins its fall, as given by \begin{equation}\label{mvR}
m_{vR}=\frac{m_o}{\sqrt{1-v_R^2/c^2}}.
\end{equation}

From the reference frame at $r$, the mass of the body at $R$ will be increased above its rest mass $m_{Rr}$ in the same proportion.  That is the value of the body's mass that remains constant during the fall from $R$ to $r$, and will be measured as $m_v$ when it arrives at $r$.  Using the equation of motion (\ref{rdot2}) and other equations previously derived, it can be shown that, regardless of any initial velocity imparted to the body at $R$, the mass/velocity relationship of (\ref{mv}) applies when the body arrives at $r$.

\section{Gravitational redshift}

For static fields there are two distinct physical explanations for gravitational redshift that can be found in the literature.  The most common view is that the redshift is caused by a loss (gain) in the energy of a photon as it climbs (descends) in a gravitational field.  Another view, argued by Okun, Selivanov and Telegdi \cite{okun00, okunseltel00} and noted by others (e.g., \cite{feynlectgrav, rhayward67}) is that the redshift is caused by a change in the energy levels of lifted atoms.  These viewpoints are incompatible;  if both were true the redshift would be doubled.

The concept of mass reference frames supports the view that gravitational redshift is caused by a change in the energy levels (masses) of lifted atoms.  The energy of a photon would not change while traveling along a null geodesic in a static gravitational field \cite{vera81, savickas94}.  If a photon is emitted at elevation $O$ with proper energy $E_o$, it will arrive at a higher elevation $x$ with the same energy. However, all stationary masses and energies at $x$ are increased relative to $O$.  To an observer at $x$, the photon arrives with a reduced energy compared to a photon at $x$ emitted by the same process.  Using a uniform field approximation of $g(x)\approx g_o$ and $E=hf$, it can be shown that a photon with proper energy $E_o$ emitted at $O$ arrives at $x$ with a frequency shift of  \begin{equation}\label{redshift}
 \frac{\Delta f}{f} = - \frac{g_o x}{c^2},
 \end{equation} relative to the frequency the photon would have if it were emitted at $x$ with proper energy $E_o$.  The proper frequency of the photon at $x$ is established by its proper energy there, in combination with Planck's constant, and is not an inherent property of the photon.  Experimental evidence of the local position invariance of Planck's constant \cite{kentmohag12} is consistent with this interpretation.

 A review of gravitational redshift experiments (e.g., \cite{pound64}) indicates that they are unable to distinguish whether photons lose energy during travel, or appear to be redshifted due to changes in mass/energy reference frames.  The proposed interpretation also satisfies the thought-experiment of Nordtvedt on the conservation of energy in gravitational redshift experiments \cite{nordtvedt75}.  The proposed existence of mass reference frames appears to be consistent with predictions and observations of gravitational redshift.

 This discussion applies only to static fields.  The time-varying field created by the expansion of the universe causes a cosmological redshift that is distinct from that of static fields.

\section{Other metrics for a central mass}

Consistent with Birkhoff's law \cite{birkhoff23}, many metrics describe the field of a central mass, each related to the Schwarzschild solution by a coordinate transformation.  Two such metrics are the isotropic and harmonic metrics \cite{bodwill03}. The isotropic metric for a central mass can be written in the form \begin{equation}\label{isometric}
d s^2=c^2 \beta_{r\infty}^2 d t^2 -\alpha^2 (d r_I^2+r_I^2 d \theta^2 +r_I^2 \sin^2 \theta d \phi^2),
\end{equation} where \begin{equation}\label{biso}
\beta_{r\infty} = (1-GM/2c^2 r_I)/\alpha,
\end{equation} and \begin{equation}\label{isoalpha}
\alpha = 1+GM/2c^2 r_I.
\end{equation}

If a body is lifted from $r_I$ to $R$ in the isotropic metric, and then allowed to fall back to $r_I$, the mass/velocity relationship of (\ref{mv}) is found at $r_I$.

If this exercise is repeated for the harmonic metric \cite{bodwill03}, that result is found once again.  What this confirms is that mass reference frames are independent of the choice of coordinates, as should be expected.  It can also be shown that at the same point in space, as determined by well-known coordinate transformations between the three metrics (e.g., \cite{bodwill03}), both $\beta_{r\infty}$ and the local gravitational acceleration $g(r)$ are identical for all three metrics.

What this suggests is that, in static fields, gravitational energy is stored as part of the masses of bodies in accordance with the simple scalar function $\beta_{ab}$, which transforms in a well-behaved way and is independent of the choice of coordinates.

\section{Non-radial motion}
Up to this point, only radial motion relative to a central mass has been considered.  Next it is shown that motion in any oblique direction is not only consistent with the concept of mass reference frames, but corresponds to the Schwarzschild metric.

Consider a stationary observer fixed at some radius $r$ from a central mass.  To that observer, special relativity and (\ref{mv}) indicate that a moving or orbiting body at $r$ should have a mass $m_{vrr}$ of \begin{equation}\label{mvrr}
m_{vrr}=\frac{m_o}{\sqrt{1-v_r^2/c^2}},
\end{equation} where $m_{vrr}$ denotes relativistic mass $m_v$ of the body at $r$ in the reference frame at $r$.  From a reference frame at infinity, the mass of the moving body at $r$ appears smaller, since all masses at $r$ appear decreased relative to those at infinity by a factor $\beta_{r\infty}$, from (\ref{mrinf}).  From that frame, the mass of the moving body at $r$ is perceived to be \begin{equation}\label{mvrinf}
m_{vr\infty}=\frac{\beta m_o}{\sqrt{1-v_r^2/c^2}},
\end{equation} where the subscripts on $\beta_{r\infty}$ are dropped for now.  If the body is in free-fall around the central mass, moving in any direction with no other external forces, its total mass will remain constant during its fall or orbit. Therefore, the body's total mass from the reference frame at infinity is constant as follows: \begin{equation}\label{mvK}
\frac{m_{vr\infty}}{m_o} = K = \frac{\beta}{\sqrt{1-v_r^2/c^2}},
\end{equation} where $K$ is a constant of the motion.  This formula makes it easy to find the velocity at any radius $r$.  If $K>1$ the body has sufficient relativistic mass \--- and associated velocity \--- to escape the central mass.  If $K<1$ the orbit will be bound.  This result is similar to that of \cite{augousti11}.

Using the expressions $v_r \equiv d x/d t_r$ and $d t_r=\beta d t$, (\ref{mvK}) can be rearranged to yield \begin{equation}\label{metK}
\frac{c^2 \beta^4 d t^2}{K^2}=c^2\beta^2 d t^2 - d x^2,
\end{equation} where $d t$ is the time of a clock at infinity.  Let $d \tau$ represent the time of a clock on the moving body.  Those two times are related by \begin{equation}\label{dtaudt}
\frac{d \tau}{d t} = \beta \sqrt{1-v_r^2/c^2}.
\end{equation}

Using (\ref{mvK}) and (\ref{dtaudt}), it can be shown that the left hand side of (\ref{metK}) is $c^2 d \tau^2=d s^2$, so that \begin{equation}\label{ds2x}
d s^2= c^2 \beta^2 d t^2 -d x^2.
\end{equation}

As previously defined, $d x$ is proper length in the motionless reference frame at $r$.  For any given metric, $d x$ can be related to the coordinates used in that solution.  For example, for the standard form of the Schwarzschild solution \begin{equation}\label{dx2schw}
d x^2 = \frac{d r^2}{\beta^2} + r^2 d \theta^2 + r^2 \sin^2 \theta d \phi^2.
\end{equation} This can be found from the metric at a synchronous instant in time $(d t=0)$, which gives the spacelike spacetime interval $d s^2 \to d x^2$ when the appropriate sign is used.  Substituting this into (\ref{ds2x}) yields \begin{equation}\label{schagain}
d s^2 = c^2 \beta^2 d t^2 - \frac{d r^2}{\beta^2}-r^2 d \theta^2 - r^2 \sin^2 \theta d \phi^2.
\end{equation} This is the Schwarzschild metric of (\ref{ds2}).  Other metrics, like the isotropic and harmonic metrics, can similarly be obtained when the corresponding expression for $d x^2$ is used.  Although this is not an independent derivation of the metrics, it demonstrates the correspondence between mass reference frames and the metrics of GR.  Since this result is based on (\ref{mvrr}) it also indicates that a body moving in any oblique direction will exhibit the mass/velocity relationship of special relativity in a local frame.

\section{Final supporting argument for static fields}
The derivation of (\ref{mv}) confirms that mass and energy are conserved during a lift and fall cycle under the proposed interpretation.  That provides a compelling argument to support the existence of mass reference frames:  As shown below, the prevailing interpretation of gravitational energy in GR does not conserve energy and mass during a lift and fall.

Consider a body being slowly lifted in a Schwarzschild field.  The mass equivalent of the proper lift energy expended over a small distance $d x$ is given by  (\ref{dmxx}).  If total mass is indeed invariant between all elevations, then $m_{xo}=m_{o}$, contrary to (\ref{dmxo}).  Then the mass equivalent of the energy used to lift a body from $O$ to $x$ would be given by  \begin{equation}\label{Deltam}
\Delta m = \int_o^x \frac{m_o g(x) d x}{c^2}.
\end{equation} If this is integrated in a Schwarzschild field, the total mass equivalent of the energy expended while lifting the body from $r$ to $R$ would be given by \begin{equation}\label{delm}
\Delta m = m_o \ln \beta_{Rr}.
\end{equation} In weak fields, $m_o + \Delta m$ approaches $m_{Rr}$ in (\ref{mRrb}).

When the body falls from $R$ to $r$ it must arrive with a relativistic mass given by (\ref{mv}) in the reference frame of a stationary observer at $r$.  It is easy to show that the lift energy/mass and the additional relativistic energy/mass acquired by the body when it returns to $r$ are not conserved, as follows: \begin{equation}\label{mloss}
\Delta m < \frac{m_o}{\sqrt{1-v_r^2/c^2}} - m_o.
\end{equation} It takes less energy/mass to lift a body than arrives back at the starting point.  Thus, without the use of mass reference frames, energy and its mass equivalent are not conserved within a lift/fall cycle.  This does not indicate any flaw in GR, but suggests that mass reference frames are necessary to complement the theory.

\section{Mach's principle}
To the extent that distant stars contribute to the local gravitational potential at any point, the proposed interpretation implies that the magnitude of the inertia of any body is directly influenced by masses throughout the universe.  Thus, the proposed interpretation brings GR into closer harmony with Mach's principle.  Nevertheless, the concepts developed herein are more likely to fuel the recurring debate on Mach's principle than to resolve it.

\section{Time-varying fields}
This paper focuses on static fields, which encompass the most important tests of GR.  Nevertheless, it is important to show how the proposed interpretation is limited in time-varying fields.  Such fields introduce issues that are discussed only briefly here.

The analysis of time-varying fields begins with the simple thought-experiment shown in Fig. 2.  Consider two masses $M_1 = M_2$ located deep in intergalactic space and separated by distance $d$.  Initially, both masses are approximately co-moving with the cosmological fluid.  The attractive gravitational force between the bodies is $F$.  Imagine that a long cable is attached to $M_2$, with the other end attached to a winch W secured to a distant, massive object.  The winch pulls on the cable with a tension of $T = 2F$, so that mass $M_2$ moves away from $M_1$ at the same rate that $M_1$ falls towards $M_2$.  Both masses will accelerate towards $W$ while maintaining distance $d$ between them.  Once they have achieved some large velocity $v$, let mass $M_2$ be pulled a great distance away from $M_1$, leaving $M_1$ moving at velocity $v$ in empty space.

\begin{figure}[htb]
\begin{center}
\epsfig{file=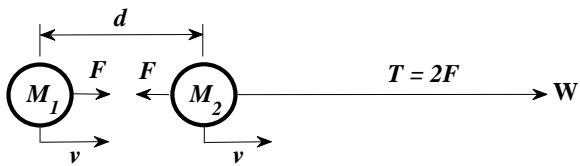, width=8.5 cm, angle = 0}
\caption{Two masses $M_1 = M_2$ in empty space, with a mutual gravitational force $F$.  A long cable pulls on $M_2$ with a tension $T \approx 2F$, causing both masses to accelerate to the right at the same rate, maintaining their separation $d$. After some time, $M_2$ is pulled away from $M_1$, leaving $M_1$ with final velocity $v$.  Energy has traveled from the winch W to $M_2$, and then through empty space to $M_1$.}
\end{center}
\end{figure}

In this example, body $M_1$ has been accelerated by gravity to velocity $v$, imparting relativistic energy/mass to the body. Where has that energy come from?  If Vera's interpretation of falling bodies from Fig. 1 is applied, then the total mass remains constant and the original rest mass $m_o$ has been converted into relativistic mass $m_v=m_o$.  If the body's motion is then stopped, it will lose the relativistic part of its mass and drop to a new rest mass of $m'_o<m_v$.  Then $m'_o < m_o$ in a region where the mass reference frame is unchanged.  The body's rest mass has been decreased by the motion induced in Fig. 2.  Conversely, the rest mass of $M_2$ would necessarily increase. That interpretation would require the rest masses of all bodies and elementary particles to vary with their prior motion in time-varying gravitational fields.  In the absence of experimental evidence of an appreciable variability in the rest masses of protons or electrons, it is evident that Vera's interpretation of a falling body does not work in a time-varying field.

Fortunately, all is not lost.  The most essential of Vera's concepts is that gravitational energy is stored as part of the masses of bodies rather than in empty fields.  In Fig. 2, the energy expended by the winch travels directly to $M_2$ via the cable.  Half of that energy then travels through space almost immediately to $M_1$, limited only by $c$.  No energy would be stored in fields over the long-term.

What can be deduced from this exercise is that gravitational energy must travel through space in a time-varying field.  Bondi described the movement of energy across the empty space between two bodies \cite{bondi90}.  In static fields, energy may be exchanged between gravitating masses, but the net change in the mass of a test particle during free-fall should be zero.  Thus the proposed interpretation of gravitational energy is compatible with time-varying fields, but is limited accordingly.  This approach is consistent with the transport of energy by gravitational waves, and also suggests that gravitons, if they exist, should convey mass/energy.  The transfer of energy in Fig. 2 is fundamental and does not arise from the concept of mass reference frames.

To show how the proposed interpretation works in time-varying fields, it is next applied to the most important cosmological metric of GR \--- the Robertson-Walker metric, which can be written in the following form:  \begin{equation}\label{RWmetric}
d s^2=c^2 d t^2-a^2(t)[d \chi^2+\sin^2 \chi (d \theta^2 + \sin^2 \theta d \phi^2)],
\end{equation} where $t$ is the cosmological time of all observers co-moving with the cosmological fluid, $a(t)$ is the scale factor representing the expansion of the Universe, and $d \chi$ represents radial coordinate distance.  Proper distance $d x$ in the radial direction is related to $d \chi$ by $d x = a(t) d \chi$ at any synchronous time $t$.

Suppose that at some time $t_1$ a particle is given an initial proper radial velocity $V$ relative to the cosmological fluid,.  At some later time $t$, it will arrive at a second point with proper velocity $v_r \equiv d x / d t$.  From the geodesic equations of motion (e.g., \cite{hobson07}), the proper radial velocity of the body at $t$ can be found:  \begin{equation}\label{vr}
v_r(t) = \frac{d x}{d t} = \frac{V}{\sqrt{\frac{V^2}{c^2}+ \frac{a^2(t)}{a^2(t_1)} \left( 1- \frac{V^2}{c^2} \right)}}.
\end{equation}  As expected, the equation confirms that if a particle starts out with an initial velocity $V=0$ relative to the cosmological fluid it will continue to move with the cosmological fluid without relative acceleration.  Moreover, a photon with an initial proper velocity of $V=c$ will maintain a proper velocity of $c$.

As previously discussed for the Schwarzschild solution, suppose a particle is slowly ``lifted" between two points.  What (\ref{vr}) indicates is that no matter how small the initial velocity $V>0$, the particle's proper velocity will never subsequently fall to zero. It will forever continue to drift slowly across the Universe.  This means that all points in the expanding  universe have the same gravitational potential.  There is no gravitational force or pseudo-force to lift against when slowly moving a particle. Therefore, all observers co-moving with the cosmological fluid experience the same mass reference frame.

However, (\ref{vr}) shows that as the universe expands, the proper velocity of of the particle decreases monotonically below its initial proper velocity $V$.  Therefore, its relativistic mass will decrease with time, relative to cosmological observers.  Using (\ref{vr}) the relativistic mass of a particle as a function of time is  \begin{equation}\label{mt}
m_v(t)=m_o \sqrt{1+ \frac{a^2(t_1)}{a^2(t)} \frac{V^2}{(c^2-V^2)}},
\end{equation} where $m_o$ is its rest mass.  Thus the relativistic mass decreases with time.  The change in relativistic mass with time can be found from (\ref{vr}) and (\ref{mt})  \begin{equation}\label{dmvdt}
\frac{d m_v}{d t} = - m_v \frac{v_r^2}{c^2} \frac{\dot{a}(t)}{a(t)}.
\end{equation}  This loss of relativistic mass is directly proportional to the rate of expansion of the Universe.  In a static universe, $\dot{a}(t) = 0$ and there is no loss of mass.  This result is consistent with Vera's interpretation that the mass of a falling body remains constant in a static field, while requiring the transfer of mass in a time-varying field.  It is important to recognize that the loss of mass in (\ref{dmvdt}) does not arise from the concept of mass reference frames, but is a direct consequence of the Robertson-Walker metric and is implicit in GR.

So what happens to the relativistic mass lost by a body moving relative to the cosmological fluid? The concepts developed herein suggest that mass/energy must travel through space to other bodies, most likely at the speed of light.  For the Robertson-Walker metric, the receiving bodies would be the masses comprising the expanding Universe.

Similarly, this approach can be applied to any metric to quantify the mass/energy lost or gained by falling bodies.  The concepts developed herein only begin to explore this topic.

Equation (\ref{mt}) provides another way to calculate cosmological redshift. If $m_V$ is the relativistic mass of a particle at time $t_1$ when it has initial velocity $V$, then the limit of $m_v(t)/m_V$ as $V \to c$ is $a(t_1)/a(t)$, which in terms of energy corresponds to the expected cosmological redshift of a photon \cite{hobson07}. \newline

\section{Conclusions}
The proposed interpretation of gravitational energy provides a means to explore the disposition of gravitational energy in solutions to the field equations.  The localization of gravitational energy in static fields is explained by simple concepts in which the energy is localized with the masses of bodies.  Observers at different gravitational potentials would experience different mass reference frames.  With this approach, it is shown that the relativistic mass of a falling body can be calculated directly from GR.  Thus, the correspondence between general and special relativity is enhanced.  Unlike the pseudotensor $t^{\mu\nu}$, this interpretation is independent of the choice of coordinates.  The use of mass reference frames also conserves mass and energy during a lift/fall cycle, unlike the concept of invariant mass.  Inasmuch as there currently exists no widely-accepted treatment of localized gravitational energy in GR, it is hoped that the proposed interpretation will be considered worthy of further study and discussion.  Much work remains to be done.

Many attempts have been made to incorporate a scalar field into gravitational theory. Examples are the scalar-tensor theory of Brans and Dicke \cite{brans61, barraco94} and the dilaton field \cite{fujii03}. What this paper suggests is that, in static fields, GR already includes an intrinsic scalar field $\beta_{ab}$ that accounts for the energy of the gravitational field.

Unlike static fields, time-varying fields require a net exchange of energy and mass between gravitating bodies. With this interpretation, gravitational waves and/or gravitons would convey energy and mass.

In Newtonian gravity, potential energy allows total energy to be conserved when lifting a body.  The equivalence of a small amount of mass to a large amount of energy allows total energy and mass to be conserved without it, consistent with everyday experience.

\end{document}